\documentclass[singlespacing]{elsart}
\usepackage{graphicx,amssymb,amsmath,times}
\journal{NIM}
\begin{document}
\begin{frontmatter}

\title{ANN-based energy reconstruction procedure for TACTIC $\gamma$-ray telescope and its comparison with other conventional  methods}
\author{V.K.Dhar\corauthref{cor1}} \ead{veer@barc.gov.in}, \author{A.K.Tickoo, M.K.Koul, R.C.Rannot, K.K.Yadav},
\author{ P.Chandra, B.P.Dubey $^\dagger$, R.Koul}
\corauth[cor1]{Corresponding author :}

\address{Bhabha Atomic Research Centre,\\
         Astrophysical Sciences Division.\\
         $^\dagger$ Electronics and Instruments Services Division,\\
         Mumbai - 400 085, India.\\
         Tel: 91-022-25593626}


\begin{abstract}
The energy estimation procedures employed   by different  groups, for determining the energy of the primary $\gamma$-ray using  a single  atmospheric  Cherenkov  imaging telescope,  
include  methods  like  polynomial fitting  in SIZE and DISTANCE, general  least square fitting  and   look-up table based  interpolation.
A novel energy reconstruction procedure, based on the utilization of Artificial Neural Network (ANN),  has  been  developed  for  the  TACTIC  atmospheric Cherenkov imaging telescope. The procedure  uses a 3:30:1  ANN configuration with resilient backpropagation  algorithm to estimate the energy of a $\gamma$-ray like event  on the basis of its image  SIZE, DISTANCE and zenith angle.  The new ANN-based energy reconstruction method, apart from yielding  an energy resolution of  $\sim$ 26$\%$, which is comparable  to that of other single  imaging  telescopes,   has the added advantage that it considers zenith angle dependence  as well.   
Details  of the  ANN-based  energy  estimation procedure along  with  its comparative  performance  with  other  conventional  energy reconstruction methods are  presented  in the paper  and  the results  indicate  that  amongst  all the methods considered in this work, ANN  method  yields  the best results. The performance  of  the ANN-based  energy reconstruction has also been validated  by determining  the  energy spectrum of the Crab Nebula   in the energy range 1-16 TeV,  as measured by the TACTIC telescope.
\end{abstract}
\end{frontmatter}

\section{Introduction}
\label{1}
Very  high energy (VHE) $\gamma$-ray astronomy,  in the energy range  $\sim$ 100GeV- 30 TeV,  has  matured into an exciting field of research activity during the last decade [1-3].  The recent  success of the field has mainly resulted from the development of the Cherenkov  imaging  technique [4-5] which  allows  efficient separation of photon induced showers from the hadron  background.   In  this technique,  the spatial distribution of  photons in the image (called   the Cherenkov image)  is recorded  by using  a close-packed  array of fast photomultiplier tubes (also called as the imaging camera with individual  tubes  as its pixels).  By exploiting   the  subtle  differences in  the  images produced  by  $\gamma$-ray and  cosmic-ray initiated  showers,  caused by  the  physical processes responsible for  the development of these  showers in the atmosphere,   it  becomes possible to effectively  segregate the two event  species with a high degree of accuracy. The  first  success of the Cherenkov imaging  technique  was the detection of the Crab Nebula  by the Whipple  group  in 1989  [6].   Following  this landmark  discovery  a  number of new  experiments  were  set up ( e.g.  HEGRA,  CAT, Durham Mark 6), leading  to  the discovery of   more   $\gamma$-ray sources.  The   HEGRA group  were the  first  group to demonstrate   that  stereoscopic  array of  telescopes  could  improve  the   gamma/hadron  discrimination  even further.  The progress  achieved  with  new  generation  telescopes   using  either   a single  Cherenkov telescope  ( e.g.  MAGIC [7])  or stereoscopic  arrays (e.g. VERITAS [8],  HESS [9] and CANGAROO [10])  has   conclusively  demonstrated   that   the imaging  technique  allows  substantial  removal  of the cosmic ray  background  events, thereby yielding unprecedented  sensitivity in the  $\gamma$-ray energy  range of  100 GeV to 30 TeV.   It is  primarily  because  of   the  success  of these  experiments   that   the  field  of  ground-based  $\gamma$-ray astronomy  today  boasts of a  source  catalogue  of about 70 gamma-ray sources  [3] from a  variety of  celestial objects  including  supernova  remnants,  pulsar  wind  nebulae, blazars,  and  microquasars.
\par
Apart  from detecting  new   $\gamma$-ray   sources,  one of the main  aim of the Cherenkov imaging  telescopes is to reconstruct  the  energy  spectra of the  sources.  A study of the  resulting spectral energy distributions can yield  valuable  information  about  the underlying $\gamma$-ray production mechanisms and  unusual astrophysical  environment characterizing these sources.  In addition,  differences in the observed energy spectrum of several  active galactic nuclei  can  also be used to study  absorption effects at the source or in the intergalactic medium due to the interaction of $\gamma$-rays with the extragalactic background photons [11,12].   
\par
Determining  the  energy  of  primary $\gamma$-rays  is an important advantage which  endows  the  atmospheric Cherenkov technique  with calorimetric capability. While  the light intensity in an  image (also known as image SIZE), represents a key parameter for determining the energy of the primary $\gamma$-ray, one also has to  consider its dependence on the core-position  and zenith angle  for improving the  energy resolution.
Since  the  precise information of core distance  is not available  with a single  imaging  telescope,  
the energy  resolution of these  telescopes  is generally  limited  to 
$\sim$25-35$\%$ [13-15].  On the other hand,  a  stereoscopic system  allows  unambiguous  reconstruction of the  shower  geometry  including  a  direct measurement  of core distance   which   leads  to  a  significant  increase in sensitivity and energy resolution of these systems [16,17].
\par
The main aim of this  work is   to  use   an  Artificial Neural Network  (ANN) based procedure for  estimating  the energy of  $\gamma$-ray like events, recorded  by a single  imaging telescope,   on the basis of  their image  SIZE, DISTANCE and zenith angle.   Apart from  being  used  for classification (pattern recognition) purposes, ANN  has also  been applied extensively to problems like function approximation or regression analysis.  
The feasibility  of employing   ANN  for    pattern  recognition problems  in particle
physics  has   been  studied   by  a  number of  workers   including    separating  gluon from quark  jets [18] and identification  of  the decays of the Z$^\circ$  boson into b$\Bar{b}$  pairs [19].  A feed-forward  ANN  classifier,   used  by the DELPHI  collaboration,   for  separating hadronic  decays  of the   Z$^\circ$   into   c and  b quark  pairs    has   also  yielded  very  promising  results [20].  Superior  performance  of the  neural network  approach,  as  compared  to   other   multivariate  analysis  methods  including  discriminant  analysis  and  classification trees,   has   been  
reported  for  tagging   of  Z$^\circ$ $\longrightarrow$ b$\Bar{b}$ events  at LEP/SLC  [21].
The feasibility  of employing   ANN  for  energy reconstruction  has  also been  studied   for a number of other applications (which are not related to Cherenkov imaging) and we will  only refer here  to few of these. While the  Wizard collaboration  has used  it for GILDA imaging silicon calorimeter [22],   the ANN-based  
approach  has also   been  used  for  reconstruction of the energy  deposited  in the calorimetry system of the CMS  detector [23] and  the hadronic calorimeter  of  ATLAS, Tilecal [24].   
\par
Application  of   ANN to atmospheric Cherenkov imaging data,   for  distinguishing  between  $\gamma$-ray  and cosmic-ray generated  Cherenkov  events,  has  been  studied   by several  workers  [25-28].  
Promising   results  have  also  been  reported  for the  wavefront sampling  telescope  CELESTE  [29]  where   the  ANN  method  was used  for not only  discriminating   $\gamma$-ray  and cosmic ray generated  Cherenkov  events  but also   for  determining   the  primary energy  and  the location of the shower core.  A detailed  case study  comparing  different  multivariate  classification  methods  (classification trees, kernel  and  nearest-neighbour  methods, linear  discriminant  analysis, support  vector  machines, neural network  etc.)  has  also  been  performed  in [30]
using    Monte  Carlo  simulated  data  generated  for the MAGIC  telescope.     Keeping in view  the  encouraging  results   reported  in   above  cited  literature,   the  main  thrust of this  work  is  to  use  ANN   for  determining  the energy of the  $\gamma$-rays   detected by an  atmospheric  Cherenkov  imaging telescope.    While  the  basic idea of applying  ANN for  determining  the  energy of the  $\gamma$-rays,      from  a point  source,   has   already been  used  by  us  in  recent past   for determining   the  energy spectra  of   the  Crab Nebula,  Mrk-421 [31] and Mrk-501  [32],   as measured  by  the TACTIC  telescope,   the  emphasis in  this  work  will be  on  presenting   a detailed  description  of the ANN-based  energy reconstruction methodology  and its comparison  with conventional  methods.  In addition, the other two aspects which have been incorporated in the present work are the usage of more data for training (see Section 3 for more details), to achieve a lower normalized rms error, than reported in [31],  and also to check the interpolation capability of the proposed ANN method with an independant data sample.  Finally, the performance  of  the ANN-based  energy reconstruction is  validated  by  revisiting   the  energy spectrum of the Crab Nebula  in the energy range 1-16 TeV as measured  by the TACTIC telescope.
\section{ TACTIC telescope}
\label{}
The TACTIC (TeV Atmospheric Cherenkov Telescope with Imaging  Camera) $\gamma$-ray telescope  has been in operation  at Mt. Abu ( 24.6$^\circ$ N, 72.7$^\circ$ E, 1300m  asl), India, for the last several years  to  study TeV gamma ray emission  from celestial sources.  
The telescope deploys a F/1 type tracking light collector of $\sim$9.5 m$^2$ area, made up of 34 x 0.6 m diameter, front-coated spherical glass facets which have been prealigned to produce an on-axis spot of $\sim$ 0.3$^\circ$ diameter at the focal plane. 
The telescope uses  a  349-pixel  imaging camera,   with a uniform pixel resolution  of $\sim$ 0.3$^\circ$ and a $\sim$ 6$^\circ$x6$^\circ$ field-of-view,   to record  atmospheric Cherenkov events produced  by  incoming cosmic-ray particles or  gamma-ray photons. 
The innermost 121 pixels (11$\times$ 11 matrix) are used for  generating the event trigger, based on the NNP (Nearest Neighbour Pairs) topological logic [33], by demanding a signal $\geq$ 25  pe for the 2 pixels which participate in the trigger generation. The back-end signal processing hardware of the telescope is based on   NIM and CAMAC  modules developed inhouse.
The data acquisition and control system of the telescope [34] has been designed around a network of PCs running the QNX (version 4.25) real-time operating system. The triggered events are digitized by CAMAC based 12-bit Charge to Digital Converters (CDC) which have a full scale range of 600 pC. The relative  gain of the photomultiplier tubes  is monitored   by repeatedly  flashing   a blue LED, placed   at a distance of $\sim$1.5m from the camera. 
The data  acquisition and control  of the TACTIC is handled by a network of three  PCs. 
While  one PC  is used to monitor the scaler rates and control the high voltage  to the photomultipliers,  the other PC handles the acquisition of the atmospheric Cherenkov events  and LED calibration data.  These two front-end PCs,  referred to as the rate stabilization node and the data acquisition node respectively,  along with a master node form the multinode Data Acquisition and Control network of the TACTIC Imaging telescope.   All executable routines stored on the master node are spawned on to the other two front-end nodes as and when required.
The telescope has a pointing and  tracking accuracy of better than $\pm$3 arc-minutes. The tracking accuracy  is checked  on a regular basis  with so called "point runs", where an optical  star   having  its declination  close  to that of the  candidate $\gamma$-ray source is  tracked continuously  for about 5 hours.  The  point run calibration data  (corrected zenith and azimuth angle  of the telescope  when the star image is centered)  are  then incorporated in the telescope drive system software so that  appropriate corrections   can be  applied directly  in real time  while tracking  a candidate $\gamma$-ray source. 
\par
Operating at $\gamma$-ray  threshold energy  of $\sim$1.2 TeV, the telescope records a  cosmic ray event rate of $\sim$2.0 Hz at a typical zenith angle of 15$^\circ$. The telescope has a 5$\sigma$ sensitivity of detecting  the Crab Nebula in 25 hours of observation time  and has so far   detected  $\gamma$-ray emission from the Crab Nebula, Mrk 421 and Mrk 501. Details  of the instrumentation aspects of the telescope and some of the results obtained on various candidate $\gamma$-ray sources are discussed in [31-35].   

\section{ Monte Carlo simulations for  energy reconstruction of $\gamma$-rays}
\label{}
The Monte Carlo  simulation data  used  for developing  a  procedure  for energy reconstruction of $\gamma$-rays  are  based on the CORSIKA (version 5.6211) air-shower  simulation code [36]. The simulated data-base for $\gamma$-ray showers used  about 34000 showers in the energy range 0.2-20 TeV  with an impact parameter  of upto 250m. These showers have been generated at 5 different zenith angles ($\theta$= 5$^\circ$, 15$^\circ$, 25$^\circ$, 35$^\circ$ and  45$^\circ$). A data-base of  about 39000 proton initiated  showers in the energy range 0.4-40 TeV, were  used  for studying the  gamma/hadron separation capability of the telescope  and  confirming  the matching   between  experimental and simulated  image parameter  distributions.  The  incidence  angle  of the proton showers  was simulated  by  randomizing  the   arrival  direction  of the primary  in a field of view of 6$^\circ$x6$^\circ$  around  the pointing direction of the telescope.  
Wavelength dependent atmospheric absorption, the spectral response of the PMTs and the reflection coefficient of mirror facets and light cones  has  also been  taken into account while performing the simulations. The  number of photoelectrons registered by each pixel has  then  been subjected to noise injection, trigger condition check and image cleaning.   The clean Cherenkov images were characterized by calculating their standard image parameters like LENGTH, WIDTH, DISTANCE, ALPHA, SIZE and   FRAC2 [4,5].
The same simulation data base has also been used, as per the well known standard procedure, for calculating the effective area of $\gamma$-rays as a function of energy and zenith angle and,  also the $\gamma$-ray retention factors  when  Dynamic Supercuts  are applied to the simulated data. 
Both  these  inputs  are  required  for  determining  the  energy  spectrum of a source  once   a  statistically  significant  $\gamma$-ray signal  
is observed in the  data. 
\par
Keeping  in view  the  fact  that the Cherenkov  light  emitted from the electromagnetic  cascade is  to a first order approximation  proportional  to the energy of the  primary  $\gamma$-ray, the approach  followed  in  atmospheric  Cherenkov  imaging telescopes is to determine the energy  on the basis of the image SIZE.  Since the intensity of the Cherenkov  light  is a function of core distance,  which is not possible to obtain with a single  imaging  telescope, the angular distance of the image  centroid  from the camera  center  (known as the  DISTANCE parameter)  is generally used  as an approximate measure of the impact  distance.   The  energy reconstruction  procedure  with  a  single  imaging  telescope  thus  involves  using   SIZE and DISTANCE   parameters  of  the  Cherenkov  event  for determining  energy  of the  primary  $\gamma$-ray. Although  the  method  has  been  found  to  work   reasonably  well    over a   restricted  zenith angle range of  $\leq$ 30$^\circ$,  there  is a  need  to  include   zenith  angle  dependence  in  the  energy  reconstruction  procedure  for   allowing  data  collection  over  a much  wider zenith angle range.
\par
In order  to  check  the  performance  of various  energy reconstruction  procedures  for the TACTIC  telescope,  we  have  divided  the simulated data base into  two  parts  so that  one  part  could  be  used  for preparing  the  data  for obtaining  parameterized  fits ( or  training the  ANN)  and  the remaining  for  testing.  For  smoothening  event to event fluctuation, which  are inherently present in  raw  data,  we  have  first  calculated $<$SIZE$>$ and $<$DISTANCE$>$  by clubbing  together  showers of  a  particular energy in various core distance bins   with  each bin having a size of 40m.  Furthermore,  additional  selection  criteria ( viz., accepting  events  with core distance $>$30m,  SIZE$>$50pe and  DISTANCE  between 0.4 $^\circ$ to 1.4$^\circ$ )  has also been  used  while  preprocessing  the training  data  to ensure  that  the  image  is robust with  minimum  possible truncation effects.   Imposing   a  lower bound on  the core distance    helps in rejecting  the  events   where  shower to shower  fluctuations  in the light  intensity  are  expected  to  be very  large,   as  most of the light  in this region  is produced  by local penetrating  particles  whose  number  can  vary  quite widely.  The  final  training data  file  thus  consists of  a single  table with   $\sim$ 350 rows.  
Each  row  has     4 columns   with    one  column  each for  energy,  $ <$SIZE$>$, $<$DISTANCE$>$  and zenith angle.  
\begin{figure}[h]\centering
\includegraphics*[width=0.9\textwidth,angle=0,clip]{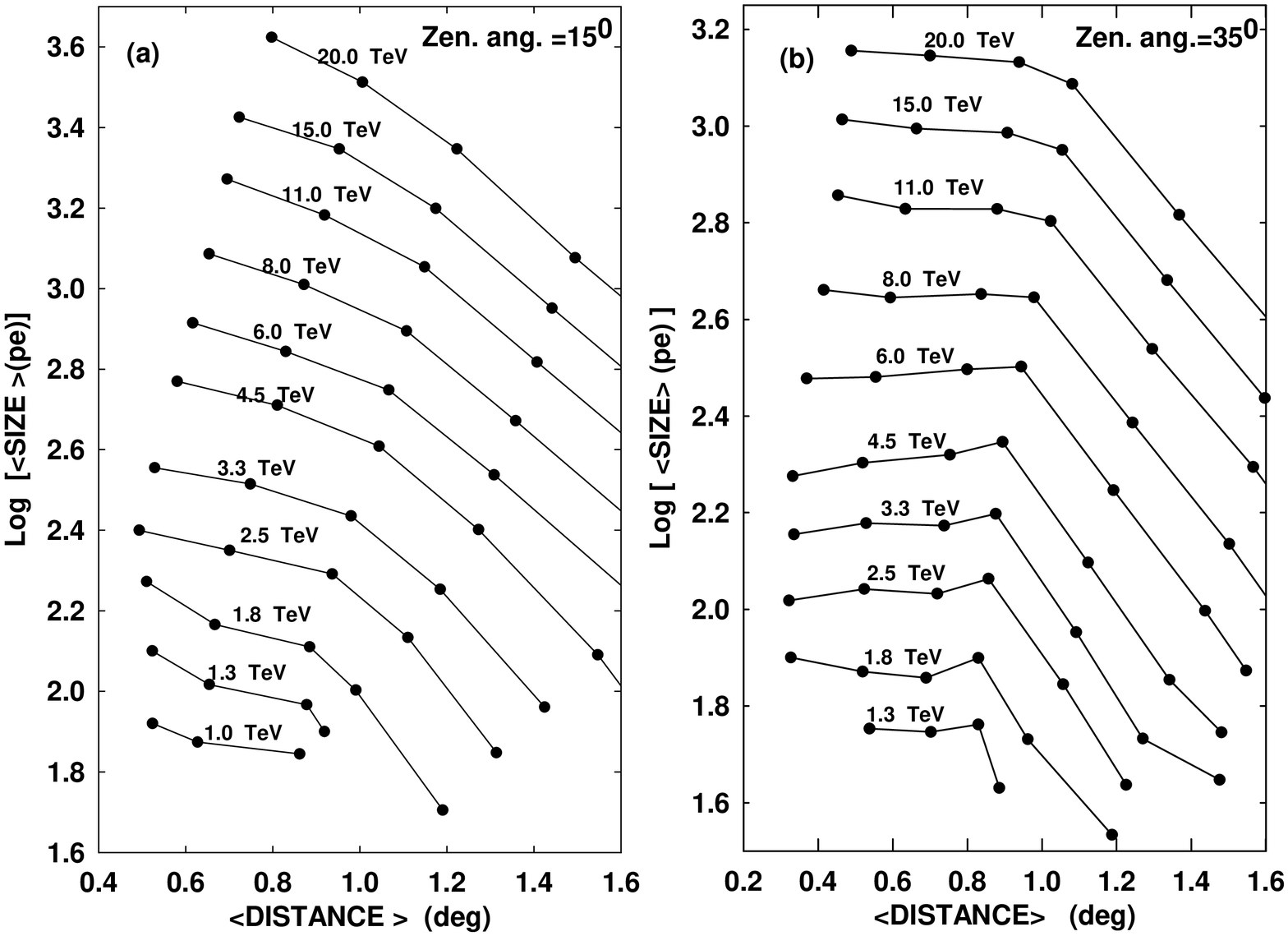}
\caption{} 
\end{figure}
It is worth mentioning   here   that  although  the  raw  data  used  for   training    is  same  as  used  in   [31],  there  is  a slight  difference  in  the  procedure   followed  for   preparing  the  training   data  file  in  this  work.  Using  the  same  number of showers   as  used in the previous  work   ( i.e  10,000),   a  new  training   data  file  of  ~350  events  ( as against   ~200  events used   by us  in our earlier work  [31])  was  generated   by   interpolating    $<$SIZE$>$   at $<$DISTANCE$>$  values  of   0.40$^\circ$, 0.50$^\circ$ ....... 1.40 $^\circ$  for   each  energy and  zenith angle.   A  representative  example   of  the   variation  of   $<$SIZE$>$   as a function of  $<$DISTANCE$>$   for   different      primary  $\gamma$-ray energies  is  shown in   Fig.1.       
It is quite evident  from this figure  that   SIZE   ( proportional  to the Cherenkov light in an image) is the most important factor which needs to considered  for estimating the energy of the primary  $\gamma$-ray.  Since,  for a fixed  $\gamma$-ray  energy,   $<$SIZE$>$   also  depends   on core distance  (proportional  to DISTANCE  parameter  of the image for a point  $\gamma$-ray source)  the  second   factor  which  needs  to be    considered is   the   DISTANCE  parameter.  On comparing   Fig.1a and Fig.1b,   which  show  the behaviour of  $<$SIZE$>$  at  zenith angles  of  15$^\circ$ and  35$^\circ$, respectively,  one finds  that the   zenith angle dependence   cannot be  ignored in situations  where a wider zenith angle  coverage  is required.   
\par
The  performance  of  a particular    energy  reconstruction  procedure   has  been  evaluated  by  calculating  the 
relative  error  in the  reconstructed energy ($\Delta_E$),  for  individual  $\gamma$-ray events  using  the  test data file.    The   relative error in the reconstructed energy  is  defined  as  (E$_{estm}$-E$_{true}$)/E$_{true}$, where  E$_{true}$  is the  true energy  and   E$_{estm}$ is the  estimated  energy   yielded  by  the  energy reconstruction  procedure.    The  mean value of ($\Delta_E$) as a function of E$_{true}$  and   energy resolution  (  $\sigma$($\Delta_E$) )  defined  as  the  root mean square  width of  the distribution of  $\Delta_E$   are   the   main  quantities   which  can  be  used  for  comparing   the performance  of   various  energy  reconstruction   procedures.  It is worth mentioning  that    the  energy resolution,  is  sometimes  estimated  by  calculating  rms  width  of  the  distribution  of  
ln(E$_{true}$/E$_{estm}$) [13,31] or ln(E$_{estm}$/E$_{true}$) [37].  The  only   reason  for  using     (E$_{estm}$-E$_{true}$)/E$_{true}$   in  this   work    as  against   ln(E$_{estm}$/E$_{true}$)   is  to   follow  a   more  standard   and  widely accepted  definition   of  energy resolution [38].  Nevertheless,  it  can be easily   shown   that  (E$_{estm}$-E$_{true}$)/E$_{true}$  $\sim$  ln(E$_{estm}$/E$_{true}$).        

\section{ Conventional  energy  reconstruction  methods }
\label{}
\subsection{Parameterized  fit  with   DISTANCE and SIZE  as  variables}
\label{}
The  first  energy  estimation  procedure  which  has  been  studied  here is  based on the  approach followed by Whipple group [13, 15],   where  ln(E$_{estm}$)  is  expressed  as  a  polynomial  in  ln(SIZE) and  DISTANCE.  The   approach  assumes   that,   for  a point    $\gamma$-ray  source,  DISTANCE   parameter of the image  provides an approximate  measure of the core distance.    The   validity  of this  assumption  has   also  been checked   for  the  TACTIC  telescope simulation  data   and the results  of the same  are  presented  in  Fig.2. 

\begin{figure}[h]\centering
\includegraphics*[width=0.9\textwidth,angle=0,clip]{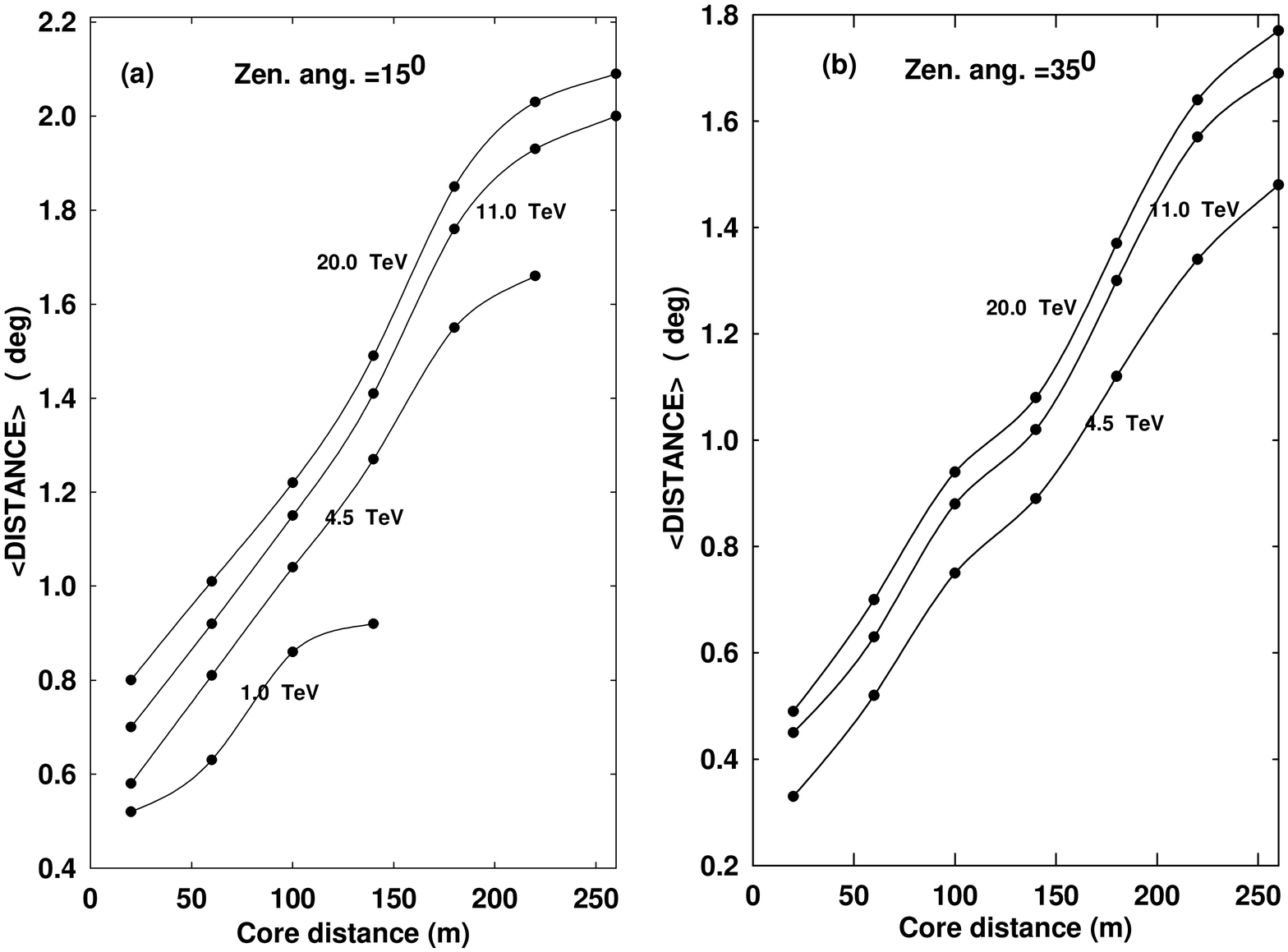}
\caption{} 
\end{figure}

The  data  used in  this  figure  has  been   generated  by clubbing  together  showers of  a  particular energy in various core distance bins  with  each bin having a size of 40m  and  finding   $<$DISTANCE$>$  for  each  core distance  bin.  Although  a  strong  correlation  between  core distance  and  $<$DISTANCE$>$    is  clearly  visible  in  Fig.2,  it  is  worth mentioning  here   that  DISTANCE   parameter of a Cherenkov  image  produced  by  an  individual  shower  is  also  dependent on the   height  of   the shower  maximum [39].   
Since for a single  telescope  it is  impossible  to  determine separately   the core distance  and  height of  shower maximum   on  an event to event basis,  obtaining   an approximate  measure of the core distance on the  basis  of  DISTANCE  parameter seems  to be the only viable solution.        
Ignoring  zenith angle  dependence and following  the  Whipple  procedure,   E$_{estm}$,   based  on  image  SIZE (S) and DISTANCE (D)  is calculated  by using  the following  relation 

\begin{equation}
  ln( E_{estm})= a_{1}+a_{2} ln(S)+a_{3} (ln(S))^2+ a_{4}( D_{0}) + a_5(D_{0}) D 
\end{equation}
Choosing $D_{0}$=1.00 $^\circ$, the values  of  $a_{1}$, $a_{2}$,  $a_{3}$, $a_{4}$($D_{0}$) and  $a_{5}$($D_{0}$),  obtained  after  fitting  equation (1)  to  the training data file  at  zenith angle of   25$^\circ$,  are found  out to be the following  : $a_{1}\sim$-2.8820, $a_{2}\sim$0.7221, $a_{3} \sim$ 0.0035, $a_{4}(D\leq$ $D_{0})\sim$ -0.2005, $a_{5}(D\leq$ $D_{0})$ $\sim$ 0.2395, 
$a_{4}(D> D_{0})$ $\sim$ -1.6766  and   $a_{5}(D > D_{0})$ $\sim$ 1.7290.  While   the  first 3 terms  in the above equation  use  the fact   that  total intensity of an image is  roughly  proportional  to the  energy of the primary,  the remaining  2 terms    modify   this  relationship   by  including  the dependence  on the core distance also.  A plot of   relative  error  in the  energy reconstruction   obtained  for test data sample  at zenith  angle 25$^\circ$  is shown in Fig.3a.  

\begin{figure}[h]\centering
\includegraphics*[width=0.9\textwidth,angle=0,clip]{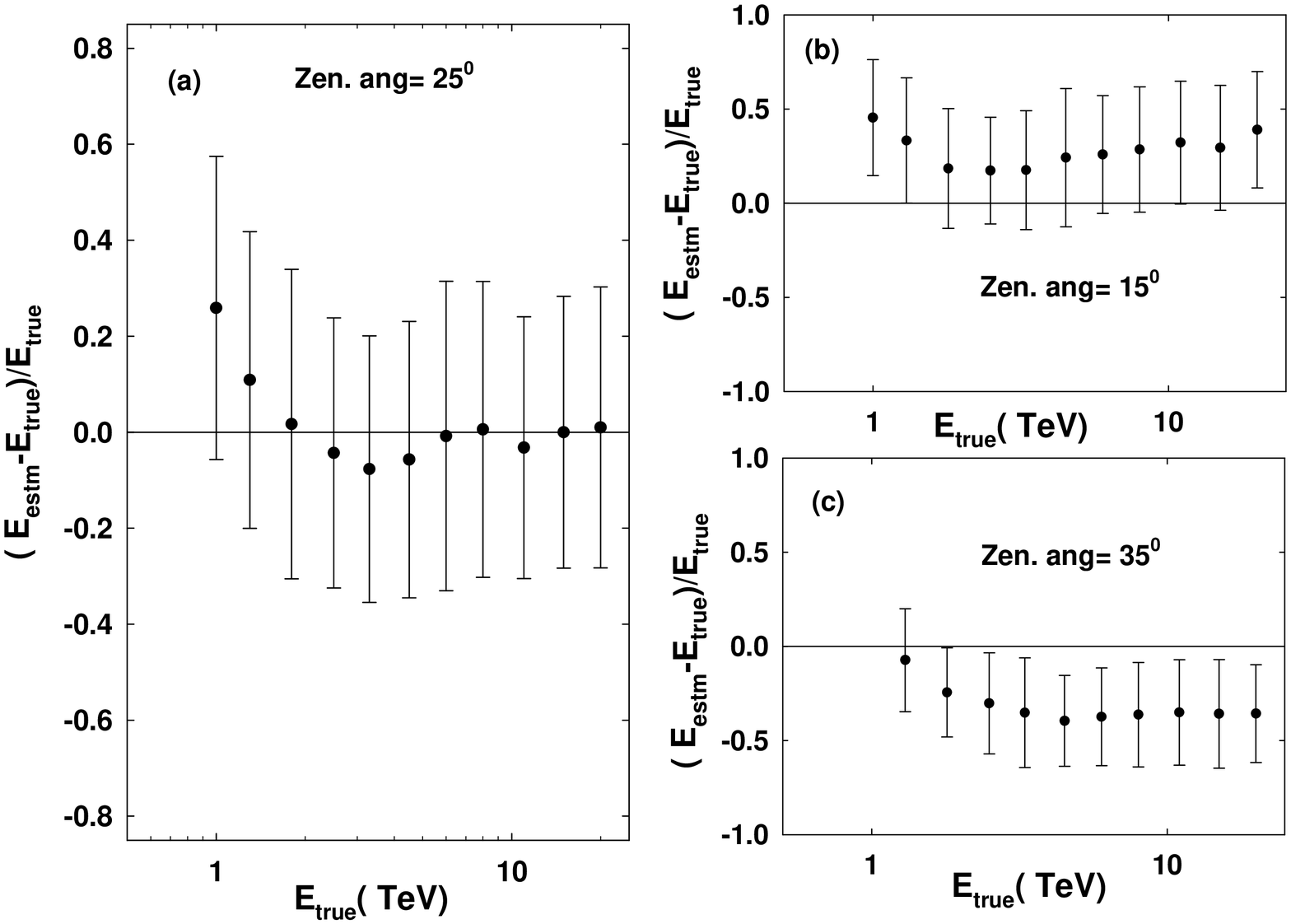}
\caption{} 
\end{figure}
The  corresponding  relative   error  in the reconstructed  energy  ($\Delta_E$)  for  zenith  angle of    15$^\circ$ and  35$^\circ$ is  also shown in (b) and (c) if  zenith  angle dependence is ignored  and  fit coefficients obtained  at   zenith angle of 25$^\circ$   are  used  as  such  in the energy   reconstruction  procedure  at  the  other two  zenith angles also.  
Although  the   energy reconstruction procedure    yields    $\sigma$($\Delta_E$)  $\sim$28$\%$  for  the  data shown in  Fig.3a,    presence  of  a  systematic  bias  seen  in   Fig.3b and  Fig.3c ($\sim$ 20 $\%$  and  $\sim$ -37 $\%$  at  zenith angles  of   15$^\circ$ and  35$^\circ$, respectively) suggests  that  there  is a need  to  include  zenith  angle  dependence  in the energy  reconstruction  procedure  for  allowing  data  collection  over  a much  wider zenith angle range.  
\subsection{ Parameterized  fit  with   DISTANCE, SIZE  and zenith angle as   variables}
\label{}
Including the zenith angle (z) dependence,  in the   energy  construction  procedure,  can  be  in principle   implemented   by  adding   one or more zenith angle  dependent  terms  to  equation (1).  The  method  followed   here  uses  guidance  from  [40]  and  employs  the following  relation  for estimating the energy.  

\begin{equation}
\ ln( E_{estm})= 1.0+ b_1 ln(S)+ b_2 \sqrt{ln(S)} + b_3 (ln(S))^2+ b_4/ cos(z) + b_5  D 
\end{equation}

The values of the  constants   after  fitting  equation (2)  to  the training data file  at  all the 5 zenith angles
(i.e   5 $^\circ$, 15$^\circ$,  25 $^\circ$,  35$^\circ$  and  45$^\circ$ ) together  are  found  out to be following : b$_1$$\sim$ 4.0053,
 b$_2$ $\sim$  -9.7814, b$_3$ $\sim$  -0.1029, 
b$_4$ $\sim$  3.3510 and  b$_5$  $\sim$  0.7822.   Plot of  relative   error  in the  estimated  energy  as a function  of    energy  for  the  test data sample at all  5 zenith  angles   is shown in Fig.4a.   

\begin{figure}[h]\centering
\includegraphics*[width=0.9\textwidth,angle=0,clip]{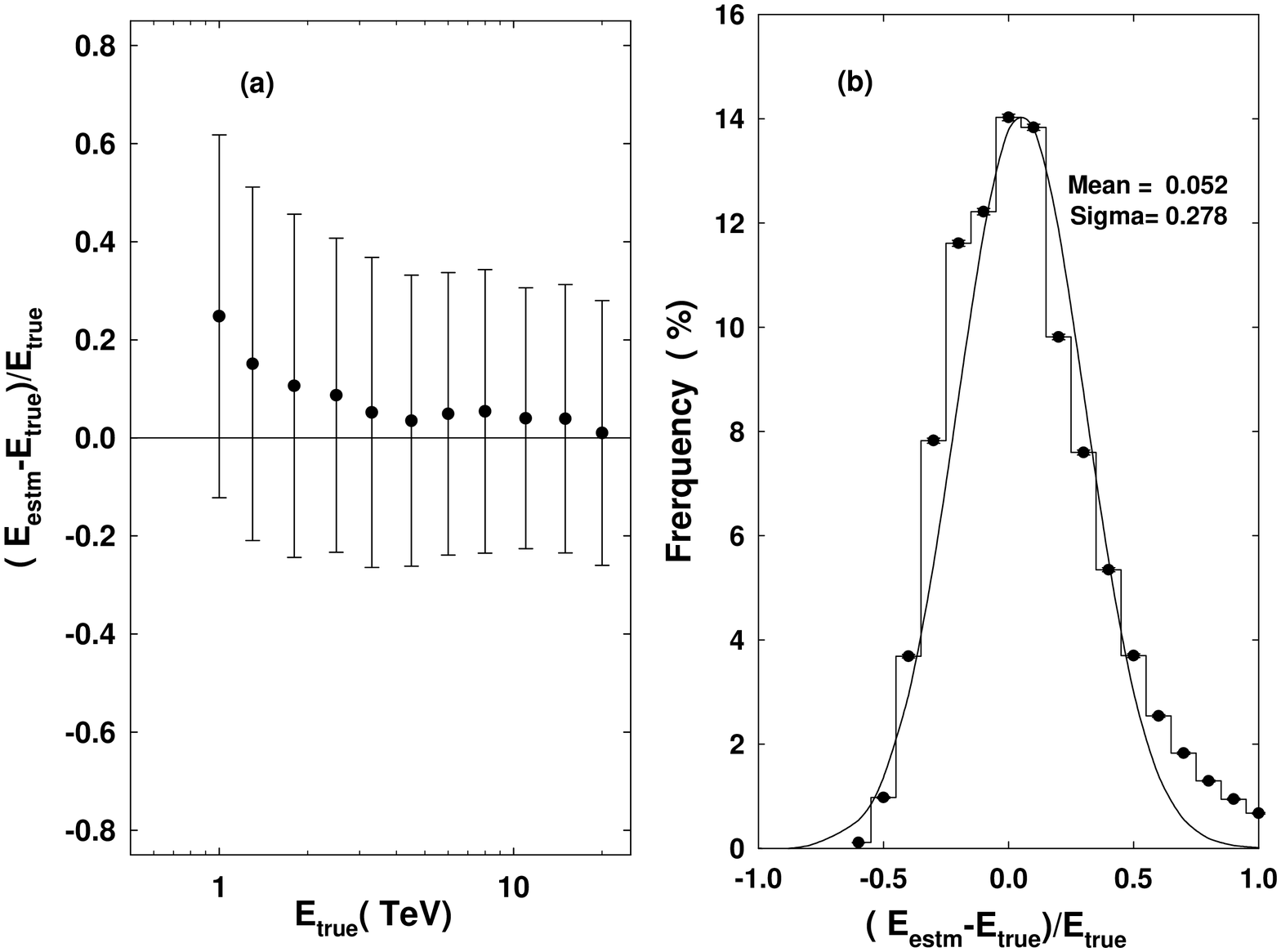}
\caption{} 
\end{figure}

Frequency distribution of $\Delta_{E}$ by  considering   almost   equal   number  showers  at  different  energies  is shown in Fig.4b.
The root mean square width of the distribution is   $\sim$  31 $\%$  and the  same  for the fitted  Gaussian  distribution is  $\sim$  28 $\%$.
Systematic  bias  at  all  energies  ($\sim$7$\%$  in the energy  range  1.8 TeV to 15.0 TeV) is also seen in Fig.4a  which  suggests  that actual  zenith angle  dependence  in the  energy  reconstruction  procedure  is   probably  more  complicated  than   what  has  been  considered  in  equation (2).   An improvement  in the  energy  resolution  has been reported  in [40] by including     an additional  parameter  called  LEAKAGE ( defined as the ratio of light content in the edge pixels to  total  light content or SIZE)  in   equation (2),  to compensate  for  leakage effects  in the relatively  small  ($\sim$  3$^\circ$  diameter ) HEGRA  CT1  camera.  Since   the TACTIC  telescope  uses a   fairly large  camera ($\sim$ 6$^\circ$  diameter)  we  do not   expect  the  energy resolution to improve   if   LEAKAGE  parameter  is  also   used.  
However,   an  attempt   to  remove  the  systematic  bias  was  also  tried  by  using a nonlinear  model  with   2 more  zenith angle dependent terms  ( viz., D/cos(z) and D ln(S)/cos(z)) in  equation (2),   but  the improvement  was found to be  only  marginal. It is worth mentioning  here  that  while the method of least squares often gives optimal estimates of the unknown parameters, it is very sensitive to the presence of unusual data points in the data used to fit a model. One or two outliers can sometimes seriously skew the results of a least squares analysis.  
\subsection{ Look-up table method using interpolation in 3 dimensions}
\label{}
The  third   energy  reconstruction  method   which  has been  studied here  is  based  on the  look-up table method. This method has  been  used quite extensively by the  HEGRA collaboration [39].  Although  the  method  was  originally  developed   for the HEGRA  stereoscopic array,  we essentially  follow  the  same principle  here. In this method,  we  generate  the  fine  grid  look-up  table   by   using  the  training   data  file.  This is done by   interpolating  the  expected $<$SIZE$>$  at finer  intervals  of  DISTANCE, energy and  zenith angle.  The  total  number  of  interpolated   SIZE  values, at a particular  zenith angle,  comprise   $\sim$4000   values   with   DISTANCE parameter  ranging from  0.4 $^\circ$   to 1.4$^\circ$  and  energy values  ranging  from  $\sim$ 0.74 TeV   to  $\sim$ 20  TeV.    In order  to  perform  interpolation  in  zenith angle,   9 different   data files   are  prepared   at   5$^\circ$  interval  in  the zenith  angle  range  from 5 $^\circ$  to 45$^\circ$.  While the above  interpolated  data has  been  obtained   by  fitting  polynomial  curves  of  order 3  to the given data points,  final energy estimation  of an event, on  the  basis  of  its  SIZE, DISTANCE and  zenith angle,  uses  only linear  interpolation.   
Plot  of  the  relative  mean    error in the  reconstructed  energy ($\Delta_E$)  as a function of energy   for test data sample  at all  the  5 zenith  angles  is shown in Fig.5a. The  frequency  distribution  of  $\Delta_E$   is  shown  in Fig.5b.          

\begin{figure}[h]\centering
\includegraphics*[width=0.9\textwidth,angle=0,clip]{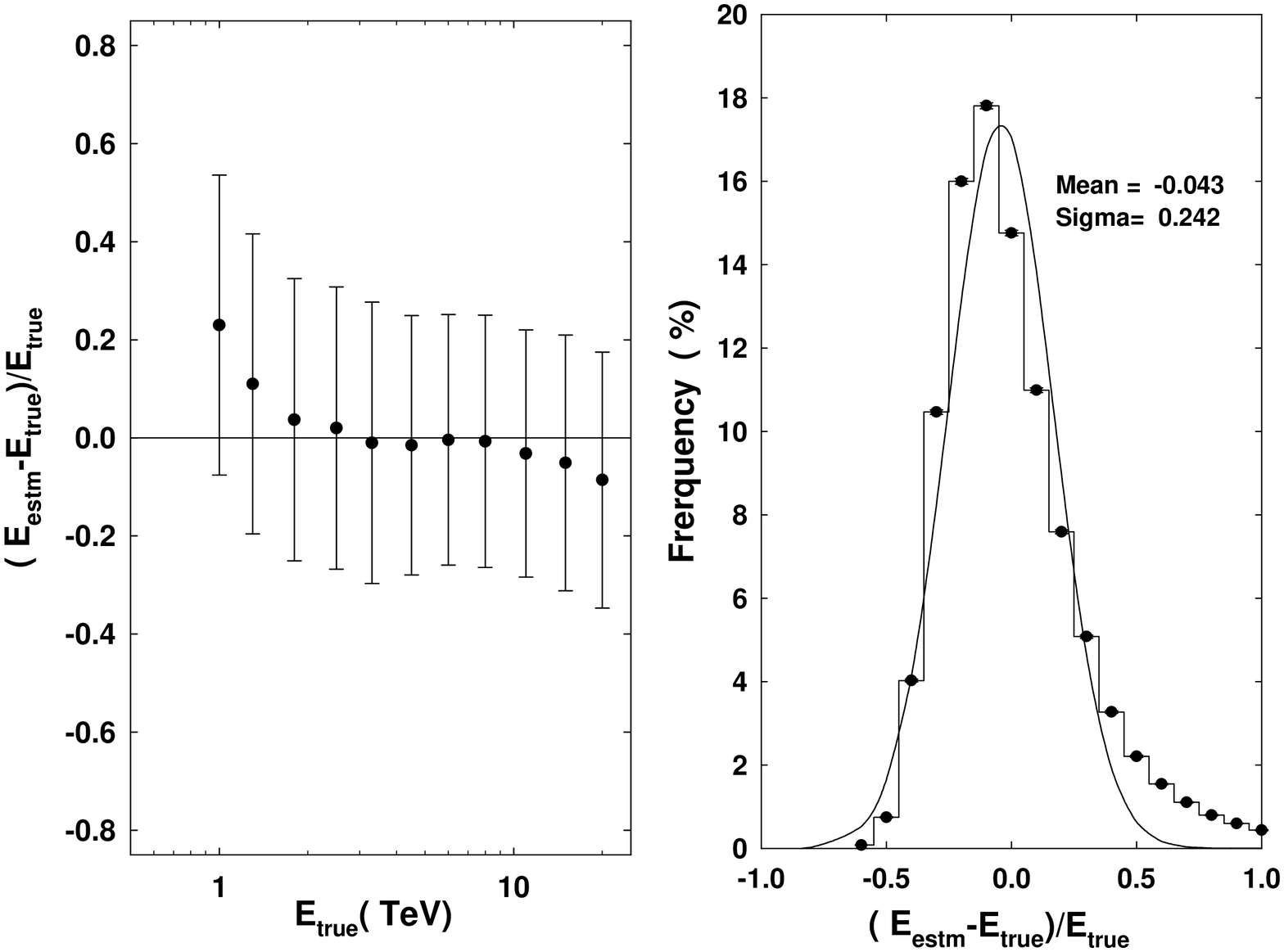}
\caption{} 
\end{figure}

It  is quite  evident  from  Fig.5a  that,   barring  energy values   at 1.0 TeV, 1.3 TeV  and  20.0 TeV  where  $|$$\Delta_E$$|$ is  found to be   $>$5.0 $\%$,  the  reconstructed   energy  has  a  negligible  bias  at   other  energy   values  from 1.8 TeV to  15.0 TeV. The rms  width of the   frequency  distribution is  found  to be  $\sim$ 27$\%$  and  the rms  width of the fitted Gaussian  distribution  is $\sim$  24$\%$.   
It is important  to  mention here  that  a positive  bias seen  in the relative  mean  error  (Fig.3a,  Fig.4a  and  Fig.5a) at energy values  of  1.0 TeV and 1.3 TeV  is  because  of the well  known   selection  effect [39]  and is  due to  sub-threshold  regime  event  triggers   because  of  their  upward  fluctuations  in the light yield  sometimes.  Since   events   with   downward   fluctuations  in the light yield  are  unable to trigger the system  in this energy range,  the energy estimates  tend to be  biased  towards larger values. 

\section{ Energy reconstruction using  Artificial Neural Network}
\label{}
\subsection{ Overview of ANN and training  of  the network }
\label{}

An artificial neural network (ANN) is an interconnected group of artificial neurons that uses a mathematical model for information processing  to accomplish a variety of tasks. They  can be configured in various arrangements to perform a range of tasks including pattern recognition and classification.  In more practical terms, an  ANN  is a  non-linear data modeling tool which  can be used to model complex relationships between inputs and outputs or to find patterns in the data. The ability of ANN to handle 
non-linear data interactions, and their robustness in the presence of  high noise levels has encouraged 
their successful use in diverse areas of physics, biology, medicine, agriculture, computer
research and  astronomy [41].  
In a feed-forward ANN the network is constructed using layers where all nodes in a given layer are connected to all nodes in a subsequent layer. The network  requires at least two layers, an input layer and an output layer. In addition, the network can include any number of hidden layers with any number of hidden nodes in each layer. The signal from the input vector propagates through the  network layer by layer till the output layer is reached. The output vector  represents the predicted output of the ANN and has a node  for each variable that is being  predicted.  The task of training  the ANN is to find the most appropriate set of weights for each connection  which minimizes  the output error.    All  weighted-inputs are summed  at the neuron node and this summed value is then passed to a transfer (or scaling)  function. 
For  a feed-forward network   with  K input nodes   described  by the input vector (x$_{1}$, x$_{2}$,......),    one hidden  layer with  J nodes  and   I  output nodes,   the output  
F$_{i}$  is given  by the following  equation

\begin{equation}
F_{i} = g  \left[ \sum \limits_{j=1}^{J} w_{ij}g  \left(\sum \limits_{k=1}^{K} w_{jk}x_k + \theta_j\right)+\theta_i\right]
\end{equation}

where    w$_{ij}$,w$_{jk}$   are the weights, $\theta_i$, $\theta_j$   are  the thresholds   and    g($\ast$) is    the activation function.     
The  training  data sample  is repeatedly  presented to the network in a number of training cycles, and the  adjustment of the free parameters 
(w$_{ij}, $,w$_{jk}$, $\theta_i$  and $\theta_j$ ) is controlled  by the learning  rate  $\eta$.  The essence of the training process is to iteratively  reduce the error between the predicted value and the target value.   While  the choice of  using  a   particular  error function  is  problem dependent,  there  is no well  defined  rule   for  choosing    the most suitable  error function. We have used  the  normalized root-mean-squared error $S_{rms}$ [42]  in this  work  which is defined as : 

\begin{equation}                   
S_{rms} =\frac{1}{PI} \sqrt{\frac{1}{2}\sum \limits_{p=1}^{P} \sum \limits_{i=1}^{I} \left(\frac{D_{p i} - O_{p i}}{D_{p i}}\right)^2}    
\end{equation}  

where D$_{pi}$ and O$_{pi}$  are the desired and the observed values    and P  is  number  of training patterns. The error here depicts the accuracy of the neural network mapping after a number of training cycles have been implemented. 
 
\par
Given the inherent power of  Artificial Neural Network (ANN)  to effectively handle the multivariate data fitting, we have developed an ANN-based energy estimation procedure  for determining the energy of  the  primary  $\gamma$-ray  on the basis of its image SIZE, DISTANCE and zenith angle.   
The procedure followed  by us uses a 3:30:1 (i.e 3 nodes in the input layer, 30 nodes in hidden layer and 1 node in the output layer) configuration of the ANN with  resilient back propagation  training algorithm  [43]
to estimate the energy of a $\gamma$-ray  event on the basis of its image SIZE, DISTANCE and zenith angle.  The  3 nodes in the input layer correspond to  zenith angle, SIZE and DISTANCE, while the 1  node  in the output layer  represents  the expected  energy  (in TeV) of the event. As already  mentioned earlier,   the  training  data   comprises  $\sim$ 350  events   where  $<$SIZE$>$  and  $<$DISTANCE $>$   are  first obtained  at  each zenith  angle   by  clubbing together showers of a particular  energy in various  core distance bins.  Apart  from reducing  the   training  data   base,  following  this  method   also makes    ANN   training   simpler  for  achieving  the  desired  level of  convergence in a reasonable  amount  of time.  The  activation function  chosen for  the present  study  is the  sigmoid function. In order to  optimize  the  number  of  nodes  required in the hidden layer, we  also varied  the number  of the nodes  in the hidden layer  from 5 to 60  in steps  of  5.  A plot  of  the  normalised  rms  error  as  a function of number of nodes  in the hidden layer  is  shown in  Fig.6a.  

\begin{figure}[h]\centering
\includegraphics*[width=0.9\textwidth,angle=0,clip]{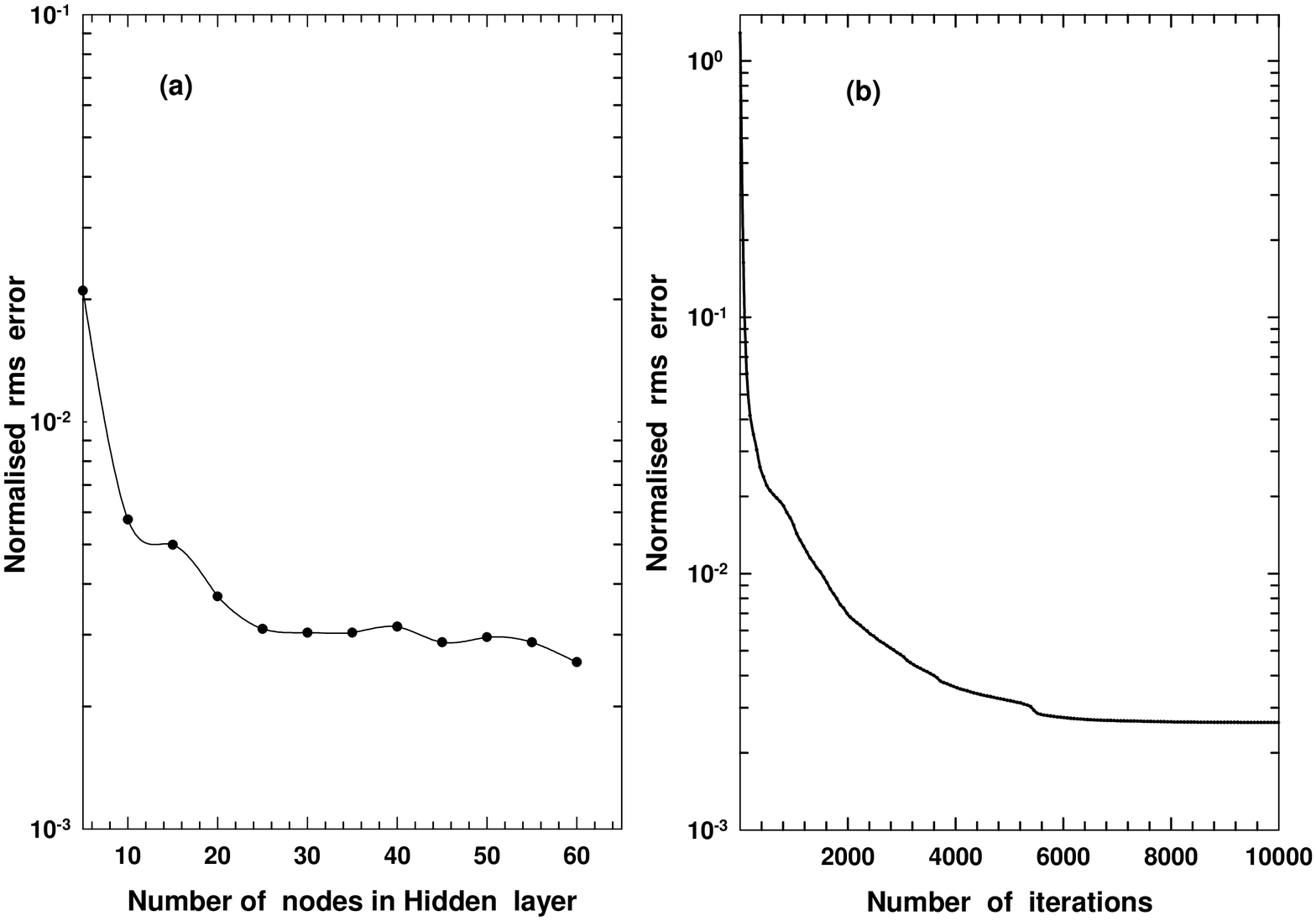}
\caption{} 
\end{figure}
Since  by   increasing  the  number  of nodes  beyond  30 results in    only   a  marginal   reduction  in the  normalised rms  error (at the cost of higher computation time),   it  seems  one  hidden layer   with  30 nodes  is  quite  optimum.  The  normalised  rms  error   at the end of the training, for  30 nodes  in the hidden  layer,    reaches  a  value  $\sim$3 $\times$ 10 $^{-3}$  and  variation  of the  same  as a function of number of iterations  is  shown  in  Fig.6b.   It  is  worth mentioning  here   that  a  normalised  rms  error  of  $\sim$2.7 $\times$ 10 $^{-2}$  was  achieved in our  previous  work  [31] and  the  improvement  seen  in  the present  work  is  as  a  result  of  using  more  data  during  ANN training.   In order  to  ensure  that  the  network    has  not  become  "over-trained" [44],   the   ANN   training  is  stopped  when  the  normalised  rms   error   stops   decreasing   any further (somewhere  around 8000  iterations).  
\subsection{ Testing and validation of the ANN}
\label{}
The  ANN  is    tested   with   two   data  samples.   The  first data sample  comprises  10,000  gamma-ray  images  (which  was earlier used for  calculating    mean SIZE and mean DISTANCE  while  preparing  the training  data set).  The   second  data  sample  comprises  24,000  gamma-ray images which  were  not   used  at all  while  preparing the  training  data set.   Both   these   data   samples    yielded   similar    energy  reconstruction  error plots,  thus  indicating   that   ANN    has  "learned"  and     not   "remembered"   the energy reconstruction procedure through  over-training [44].  Plot of energy reconstruction  error obtained  for  second test data sample  is  shown in Fig.7a. The  frequency  distribution  of  $\Delta_E$   is  shown  in Fig.7b.          

\begin{figure}[h]\centering
\includegraphics*[width=0.9\textwidth,angle=0,clip]{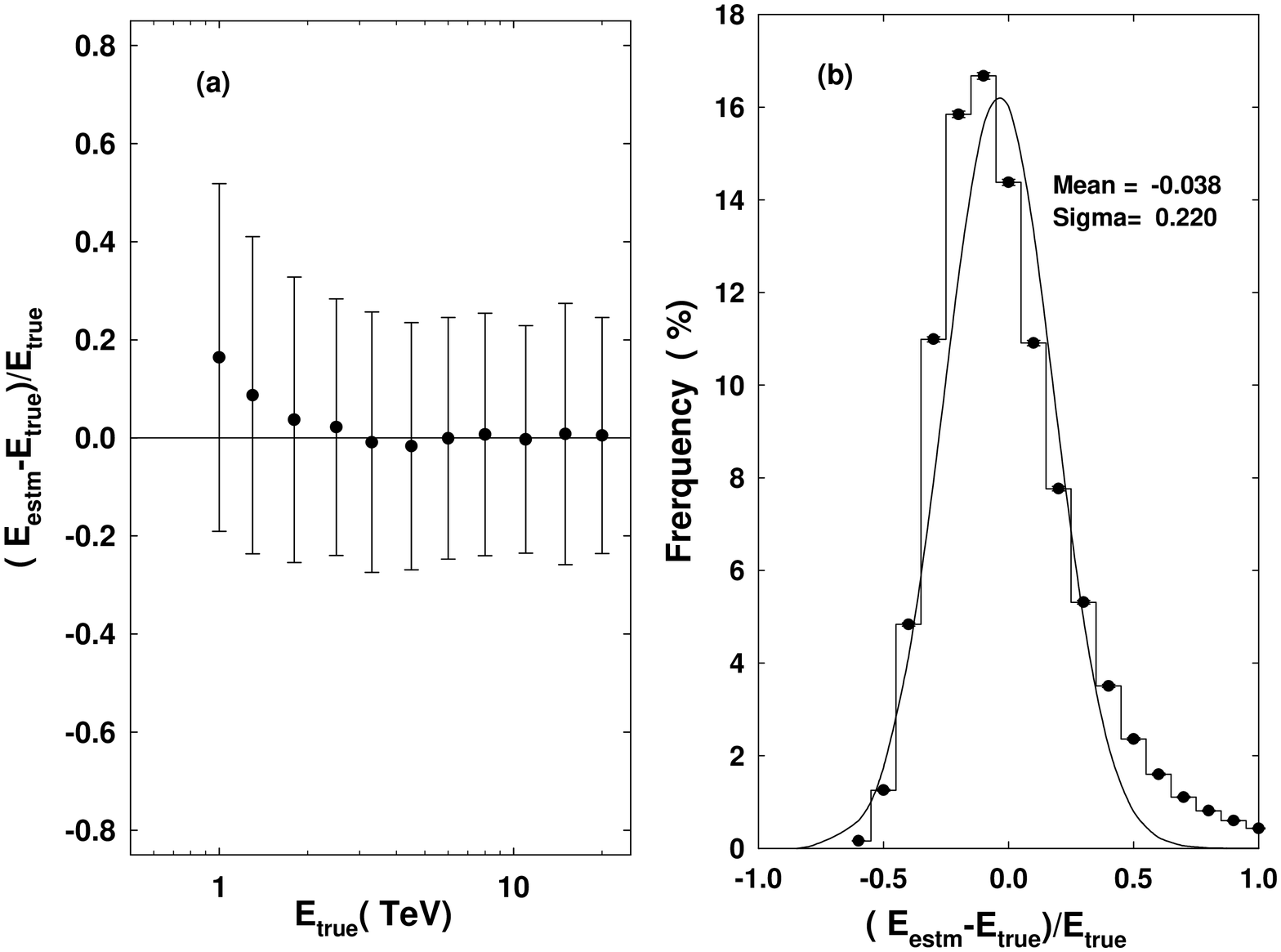}
\caption{} 
\end{figure}


It  is  evident  from  Fig.7a  that,  the  reconstructed   energy,   employing  the  ANN method,  has  a  negligible  bias   in the   energy  range  1.8 TeV to 20.0 TeV  with  $|$ $\Delta_E$ $|$  $<$ 5.0 $\%$.  The  rms  width  of  the frequency  distribution is  found  to be  $\sim$26$\%$  and the 
same for  best fit Gaussian  distribution  is  $\sim$22$\%$.
\par
The  interpolation  capability  of  the ANN-based  energy reconstruction  procedure, at  intermediate  $\gamma$-ray energies and zenith angles,  has also  been  checked  by   applying  it  to  an  independent  validation data  base  of  ~ 4000 showers.  The energy of the  primary $\gamma$-rays  was  chosen  to be   1.1 TeV,  2.1 TeV,  5.2 TeV and 9.5 TeV   at  zenith angles of 10$^\circ$  and 20$^\circ$,  and  2.1 TeV,  5.2 TeV,  9.5 TeV  and 17.0 TeV  at  zenith angles of 30$^\circ$  and 40$^\circ$. 
Since  no   simulated  data  at  these  zenith angles and energies  was  used  during  training of the ANN,  the results  obtained  now   
on the validation data   obviously  indicate   the  interpolation capability  of the  ANN. A plot of energy reconstruction  error obtained  for the validation  data  sample  is  shown in Fig.8a.  The  frequency  distribution  of  $\Delta_E$  along  with a best fit Gaussian  distribution to the histogram  is  shown  in Fig.8b. 

\begin{figure}[h]\centering
\includegraphics*[width=0.9\textwidth,angle=0,clip]{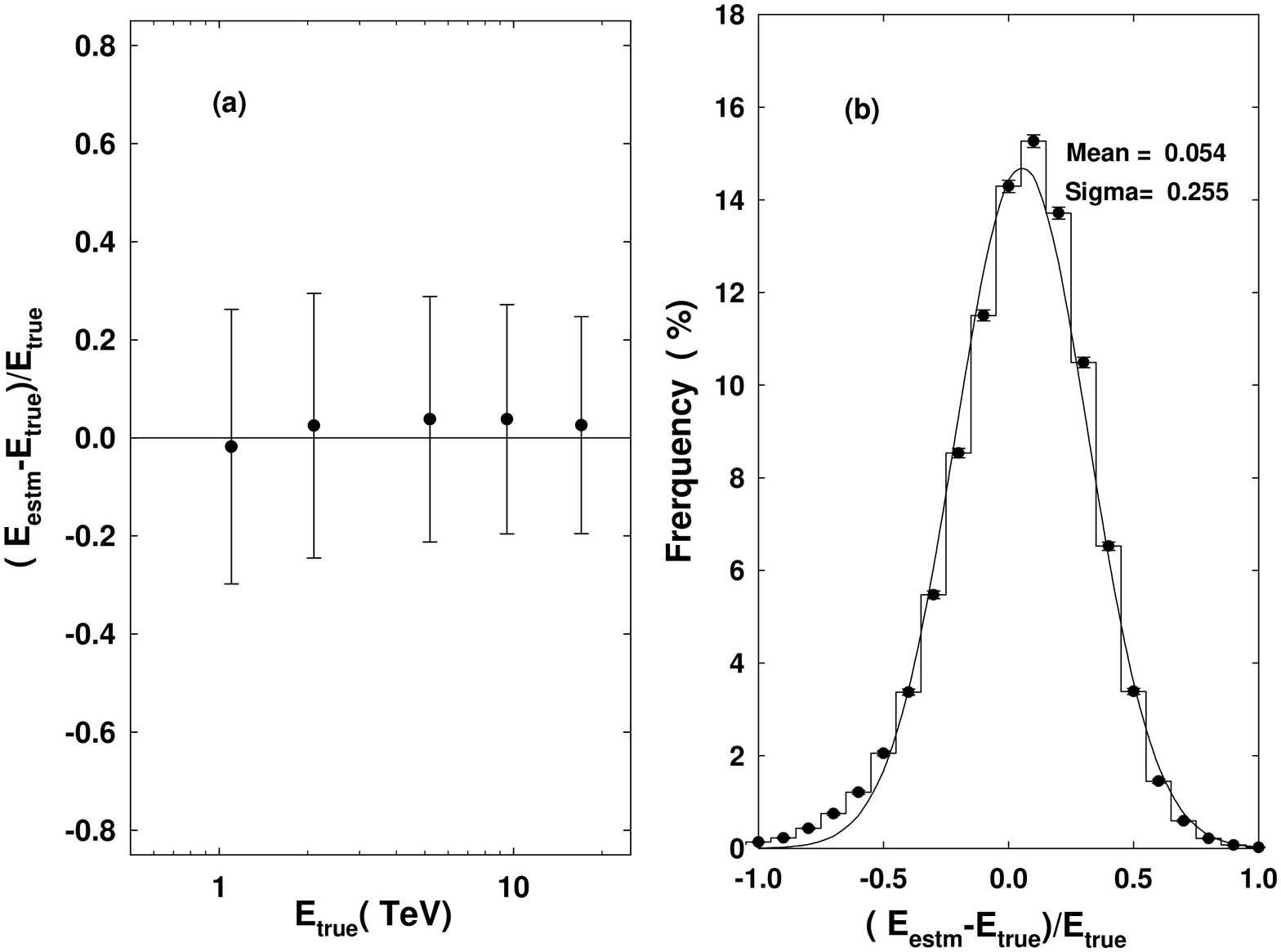}
\caption{} 
\end{figure}


The  rms  width  of  the  best fit Gaussian  distribution   for the  test and validation  data  ($\sim$ 22$\%$ and  $\sim$ 26$\%$, respectively), 
with  a  negligible  bias   in the   energy   suggests   that  the  performance  of  the  ANN-based  method   energy  reconstruction  is quite  reliable. Taking  higher  of the  two  $\sigma$($\Delta_E$) values ( i.e  $\sigma$($\Delta_E$) $\sim$ 26$\%$)  as  a safe   value  of the   energy resolution achieved  by the   ANN-based  energy reconstruction procedure, one can  easily  conclude  that  the proposed   method, apart from yielding a comparable  performance   to  that  of  other single imaging telescopes ( e.g. $\sigma$(ln E) of $\sim$ 25$\%$  reported  by  the Whipple group [15]), has the added advantage that it considers zenith angle dependence  of SIZE and DISTANCE  parameters as well.  The procedure thus allows data collection over  a much wider zenith angle range as  against a  coverage of upto 30$^\circ$ in case the zenith angle dependence is to be ignored.  On  analyzing   figures  3, 4, 5 and 7,  it is also  obvious  that the  ANN-based  energy reconstruction  procedure   yields  better results (both  in terms of bias  and energy resolution) as compared  to the  conventional  energy  reconstruction  methods.   Even though  the  look-up table based  energy reconstruction procedure  appears  to be equally  competitive  it  suffers  from  other  drawbacks.  
Implementation of  this  method  requires  cumbersome tabulation  of  interpolated  data  at  a  number of  DISTANCE, energy  and  zenith angle  values and   energy   reconstruction   procedure  is  also  more time consuming  as compared  to the ANN  method.      
Implementation  of the  ANN-based energy  reconstruction  procedure, on the other hand, is  relatively  much  more  straignt  forward.   Once satisfactory training of the ANN is  achieved, the corresponding ANN generated weight-file can  be  easily used  by  an  appropriate  subroutine  of the  main data analysis program  for  determining  the energy  of  $\gamma$-ray  like  events.  Use of  a
dedicated   ANN software package  is  thus  necessary  only  during   the  training  of  the  ANN. 
Hence, compared  to the conventional  methods, the  ANN-based energy  reconstruction  procedure    offers  several  advantages  like   reasonably  good   energy resolution,  applicability  over  a  wider  zenith angle  range  and  implementation  ease.        

\section{ Energy spectrum  of the Crab Nebula as measured  by the TACTIC telescope }
\label{}
In order to test the validity of the ANN-based energy estimation procedure, we have  applied  this  procedure  for determining  the energy spectrum of the Crab Nebula.   For this purpose  we  reanalyzed   the   Crab Nebula  data collected by  the TACTIC  imaging telescope  for $\sim$101.44 h during  Nov. 10, 2005 - Jan. 30, 2006. The zenith angle of the observations was $\leq$45$^\circ$.  The  data has been
collected with inner 225 pixels ($\sim$ 4.5$^\circ$ $\times$ 4.5$^\circ$)  of the full imaging camera  with the innermost 121 pixels 
($\sim$ 3.4$^\circ$ $\times$ 3.4$^\circ$) participating  in the trigger.  The  data recorded  by the telescope   was corrected for inter-pixel gain variation and  then subjected to the standard two-level image  'cleaning' procedure  with picture and boundary thresholds of  6.5$\sigma$ and 3.0$\sigma$, respectively.  The clean Cherenkov images were characterized by calculating their standard image parameters like LENGTH, WIDTH, DISTANCE, ALPHA, SIZE and   FRAC2 [4,5]. Before determining the energy spectrum,  the agreement between the predictions from Monte Carlo simulations and the actual performance of the telescope  has  been  checked.  This  is done   by comparing the observed trigger rate of the telescope  with the predicted value  and  by comparing  the expected  and observed  image parameter  distributions  for  protons [35].  Reasonably good  matching is  seen  between  the experimentally  observed  quantities  and  those  predicted  by simulations.
\par
The standard   Dynamic  Supercuts  procedure  [13,31]  is   used  to separate   $\gamma$-ray like images  from  the background  cosmic rays. The Dynamic Supercuts $\gamma$-ray selection  criteria  used  in the  present  analysis  are   slightly  less tight   than the ones used by us in our  earlier work [31,32]  as the  main aim  here is  to  increase the number  of  $\gamma$-ray like events  with  only a  marginal loss  of  statistical   significance.  The  new  cuts  values used  for the present analysis  are the  following :    $0.11^\circ\leq  LENGTH \leq(0.260+0.0265 \times \ln S)^\circ $,   $0.06^\circ \leq WIDTH \leq (0.110+0.0120 \times \ln S)^\circ$,
$0.52^\circ\leq DISTANCE \leq 1.27^\circ cos^{0.88}z$, $SIZE \geq 450 d.c$ ( where 6.5 digital counts$\equiv$1.0 pe ),
$ALPHA \leq 18^\circ$  and $FRAC2 \geq 0.35$.   The number of $\gamma$-ray events  obtained after applying  the above cuts   are  determined  to be   $\sim$(928$\pm$100) with a statistical significance of $\sim$9.40$\sigma$. Defining $ALPHA\leq18^\circ$ as the $\gamma$-ray domain 
and $27^\circ\leq ALPHA\leq81^\circ$ as the background region, the number of $\gamma$-ray events have  been  calculated  by subtracting the expected number of background events (calculated  on the basis of background  region) from the  $\gamma$-ray domain events. 
\par
The differential photon flux  per energy bin  has been computed using the formula

\begin{equation}
\frac{d\Phi}{dE}(E_i)=\frac {\Delta N_i}{\Delta E_i \sum \limits_{j=1}^5 A_{i,j} \eta_{i,j} T_j}
\end{equation}

where $\Delta N_i$ and $d\Phi(E_i)/dE$ are the number of events and the differential flux at energy $E_i$, measured in the ith  energy bin $\Delta E_i$ and over the zenith angle range of 0$^\circ$-45$^\circ$, respectively. $T_j$ is the observation time in the jth zenith angle bin with corresponding energy-dependent effective area ($A_{i,j}$) and $\gamma$-ray acceptance ($\eta_{i,j}$). The 5 zenith angle bins (j=1-5) used are 0$^\circ$-10$^\circ$, 10$^\circ$-20$^\circ$, 20$^\circ$-30$^\circ$, 30$^\circ$-40$^\circ$  and 40$^\circ$-50$^\circ$ with  effective  collection area  and  $\gamma$-ray acceptance  values   available at 5$^\circ$, 15$^\circ$, 25$^\circ$, 35$^\circ$ and 45$^\circ$. The number of $\gamma$-ray events  ($\Delta N_i$)  in a particular  energy bin is  calculated  by subtracting the expected number of background events, from the  $\gamma$-ray domain events. 
The $\gamma$-ray differential  spectrum  obtained   after  using  appropriate values of  effective collection area and $\gamma$-ray acceptance  efficiency  (along with  their  energy and zenith angle dependence) is  shown in Fig.9a.    

\begin{figure}[h]\centering
\includegraphics*[width=0.9\textwidth,angle=0,clip]{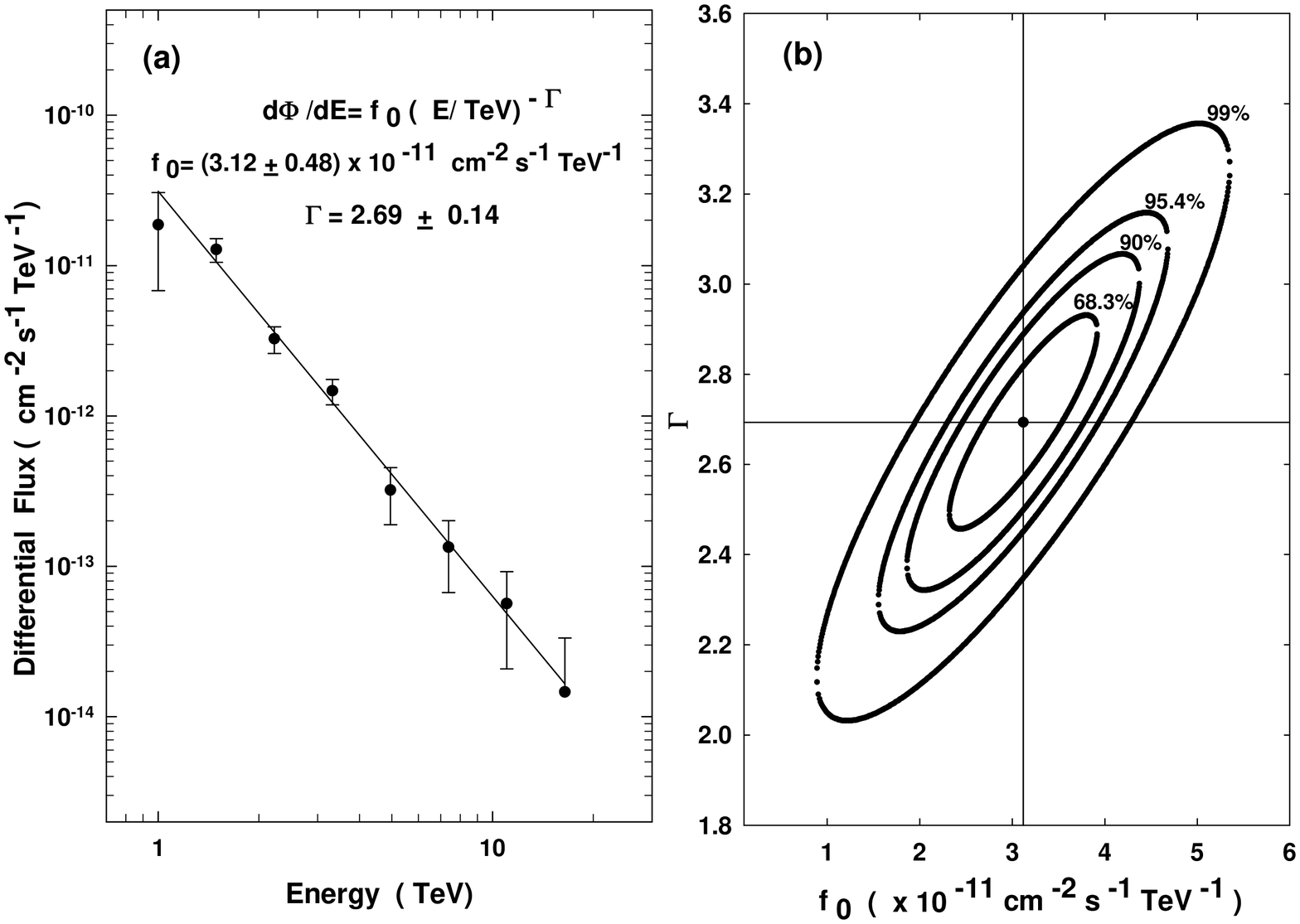}
\caption{} 
\end{figure}

A  power law  fit   $(d\Phi/dE=f_0 E^{-\Gamma})$   to the   measured  differential  flux  data  with  $f_0 \sim (3.12\pm0.48)\times 10^{-11} cm^{-2} s^{-1}TeV^{-1}$  and  $\Gamma \sim2.69\pm0.14$  is also shown in Fig.9a.  The fit  has a $\chi^2/dof\sim3.64/6$ with a corresponding probability of $\sim$0.72. The errors in the flux constant and the spectral index are standard errors. Reasonably  good  matching of this spectrum  with  that obtained  by the Whipple and  HEGRA  groups [45,46]  reassures  that the procedure
followed by us for obtaining the energy spectrum  of a $\gamma$-ray  source is quite  reliable. The  confidence ellipses in the two parameters  jointly 
(i.e  $f_0$  and  $\Gamma$)   at    68.3$\%$, 90$\%$,  95.4$\%$  and 99$\%$   confidence  levels  are  shown  in  Fig.9b. 
The  corresponding   $\Delta$$\chi^2$  values  of these  4 contours,   for  6  degrees of freedom  are   $\sim$7.04, $\sim$10.6,    $\sim$12.8  and $\sim$16.8. 
\par
It is important  to point out  here   that  for  background  cosmic ray events, which are not coming from the source   direction  and  are classified as  $\gamma$-ray like events  by the  Dynamic  Supercuts  procedure,  a wrong  energy  value   will  be  obtained  for  them.  However,   subtraction  of the background events  (estimated from  $27^\circ\leq$ ALPHA$\leq81^\circ$  region),   from   the $\gamma$-ray domain  
(defined  as $\alpha\leq18^\circ$),  will  cancel out  these events  (in a statistical sense).  Estimating the energy  spectrum  of  $\gamma$-rays  in the presence of  background  events   by  following  this  approach is well known [39] and  has  been used quite  extensively  by other groups.  


\section{ Conclusions  }
\label{}
A novel  ANN-based  energy estimation procedure,  for determining the energy spectrum of a candidate $\gamma$-ray source has  been developed. The procedure followed  by us  uses an Artificial Neural Network  to estimate the energy of a $\gamma$-ray like event on the basis of its image SIZE, DISTANCE and zenith angle.  Apart from yielding a reasonably good  $\sigma$($\Delta_E$) of $\sim$ 26$\%$, this procedure  has the added advantage that  it  allows data collection over  a much wider zenith angle range as against a  coverage of upto 30$^\circ$ only in case the zenith angle dependence is to be ignored. We have  also successfully  implemented 
the ANN-based energy  reconstruction  algorithm in our  analysis  chain,  by directly using the ANN generated weight-file,  so that the energy of a $\gamma$-ray like  event could be predicted without  using the ANN software package. Reasonably good  matching of the  Crab Nebula  spectrum as  measured  by the TACTIC   telescope with that obtained  by the Whipple and  HEGRA  groups reassures  that the procedure followed by us for obtaining the energy spectrum  of a $\gamma$-ray  source is quite  reliable.


\section{Acknowledgements}
\label{13}
The authors would like to convey their gratitude to all the concerned colleagues of the  Astrophysical Sciences  Division  for their contributions towards the instrumentation and observation aspects of the TACTIC telescope.  The  authors would also  like to thank Sh. N. Bhatt for helpful discussions related to  Section 6  of this  work.  

\newpage
\section{Figure Captions}
Fig 1. {\label{fig:1} Variation of $<$SIZE$>$  as  a function of $<$DISTANCE$>$  for   $\gamma$-rays  from  a point  source  of different  energies at  zenith angles  of  (a) 15$^\circ$ and  (b) 35$^\circ$.}

Fig 2. {\label{fig ---}  Variation of $<$DISTANCE$>$  as  a function of  core distance   for   $\gamma$-rays  of various energies  from  a point  source  at  zenith angles   of  (a) 15$^\circ$ and  (b)35$^\circ$.} 

Fig 3. {\label{fig ---} (a)  Relative  mean   error in the  reconstructed  energy ($\Delta_E$ = (E$_{estm}$-E$_{true}$)/E$_{true}$)  as a function of energy  for  zenith  angle of  25$^\circ$  using   the energy estimation procedure given by equation (1).
Relative mean   error  in the  estimated  energy  as a function  of  energy   for  zenith angles  of  (b) 15$^\circ$ and  (c) 35$^\circ$  if  zenith  angle dependence is ignored  in the energy reconstruction procedure.} 

Fig 4. {\label{fig ---} (a)  Relative  mean    error in the  reconstructed  energy  ($\Delta_E$ = (E$_{estm}$-E$_{true}$)/E$_{true}$)  as a function of energy  for the energy estimation procedure given by equation (2).   (b) Frequency distribution of $\Delta_E$  along  with a best fit Gaussian   distribution to the data.} 

Fig 5. {\label{fig ---} (a)  Relative  mean    error in the  reconstructed  energy  ($\Delta_E$ = (E$_{estm}$-E$_{true}$)/E$_{true}$)   as a function of energy  for the  look-up table based energy estimation procedure. (b)  Frequency distribution of $\Delta_E$ along  with a best fit Gaussian  distribution to the histogram.}
 
Fig 6. {\label{fig ---} (a)  Normalised root mean square  error  as a function of  number of nodes in the hidden layer. (b) Normalised  root mean square error as a function of number of iterations  for   30  nodes  in the  hidden layer.}   

Fig 7. {\label{fig ---} (a)  Relative  mean    error in the  reconstructed  energy  ($\Delta_E$ = (E$_{estm}$-E$_{true}$)/E$_{true}$)   as a function of energy  for the ANN- based energy estimation procedure. (b)  Frequency distribution of $\Delta_E$ along  with a best fit Gaussian  distribution to the histogram.}   

Fig 8. {\label{fig ---} (a)  Relative  mean    error in the  reconstructed  energy ($\Delta_E$ = (E$_{estm}$-E$_{true}$)/E$_{true}$)  as a function of energy  for the ANN- based energy estimation procedure when  applied to a validation  data sample. (b)  Frequency distribution of $\Delta_E$ along with a best fit Gaussian  distribution to the histogram..}   

Fig 9. {\label{fig ---}  (a) The differential  energy spectrum of the Crab Nebula  as measured by the TACTIC  telescope  and  employing  the  ANN-based  energy reconstruction procedure.(b) The  confidence ellipses in the two parameters  jointly (i.e  $f_0$  and  $\Gamma$)   at    68.3$\%$, 90$\%$,  95.4$\%$  and 99$\%$   confidence  levels.}   

\end{document}